\documentclass[12pt]{article}
\usepackage{graphicx}
\usepackage{amsmath} 
\usepackage{mcite}
\input{epsf} 
\def\ltap{\raisebox{-.6ex}{\rlap{$\,\sim\,$}} \raisebox{.4ex}{$\,<\,$}} 
\def\gtap{\raisebox{-.6ex}{\rlap{$\,\sim\,$}} \raisebox{.4ex}{$\,>\,$}}

\newcommand\as{\alpha_{\mathrm{S}}} 
 
\def\beq{\begin{equation}} 
\def\eeq{\end{equation}} 
\def\beeq{\begin{eqnarray}} 
\def\eeeq{\end{eqnarray}} 
 
\def\to{\rightarrow}

\def\WH{{$W\!H$}}
\def\ZH{{$Z\!H$}}
\def\VH{{$V\!H$}}
\def\Hbb{$H\to b{\bar b}$}
\def\MC@NLO{{\sc MC@NLO}}
\def\POWHEG{{\sc POWHEG}}
\def\PYTHIA8{{\sc PYTHIA8}}
\def\HERWIG++{{\sc HERWIG++}}

\begin{document} 

\begin{titlepage}
\begin{flushright}
DF-08-2013\\
IFUM-1019-FT\\
ZU-TH 27/13
\end{flushright}
\renewcommand{\thefootnote}{\fnsymbol{footnote}}
\par \vspace{10mm}

\begin{center}
{\Large \bf Higher-order QCD effects for associated}
\\[0.5cm]
{\Large \bf \WH\ production and decay at the LHC}
\end{center}
\par \vspace{2mm}
\begin{center}
{\bf Giancarlo Ferrera}$^{(a)}$, 
{\bf Massimiliano Grazzini}$^{(b)}$\footnote{On leave of absence from INFN, 
Sezione di Firenze, Sesto Fiorentino, Florence, Italy.}~~and~~{\bf Francesco Tramontano}$^{(c)}$\\

\vspace{5mm}

$^{(a)}$ Dipartimento di Fisica, Universit\`a di Milano and\\
INFN, Sezione di Milano, I-20133 Milan, Italy\\

$^{(b)}$ Institut f\"ur Theoretische Physik, Universit\"at Z\"urich, CH-8057 Z\"urich, Switzerland\\

$^{(c)}$ Dipartimento di Fisica, Universit\`a di Napoli ``Federico II''and\\
INFN, Sezione di Napoli, I-80125 Naples, Italy

\vspace{5mm}

\end{center}

\par \vspace{2mm}
\begin{center} {\large \bf Abstract} \end{center}
\begin{quote}
\pretolerance 10000

We consider Standard Model Higgs boson production
in association with a $W$ boson in hadron collisions.
We supplement the fully exclusive perturbative computation of QCD radiative effects
up to next-to-next-to-leading order (NNLO) 
with the computation of the decay of the Higgs boson into a $b{\bar b}$ pair
at next-to-leading order (NLO).
We consider the selection cuts that are 
typically applied in the LHC experimental analysis,
and we compare our fixed-order predictions with
the results obtained with the \MC@NLO event generator.
We find that NLO corrections to the $H\to b{\bar b}$ decay can be important to
obtain a reliable $p_T$ spectrum of the Higgs boson, but that, in the cases of interest,
their effect is well accounted for by the parton shower Monte Carlo.
NNLO corrections to the production process typically
decrease the cross section by an
amount which depends on the detail of the applied cuts,
but they have a mild effect on the shape of the Higgs $p_T$ spectrum.
We also discuss the effect of QCD radiative corrections on the invariant mass
distribution of the Higgs candidate.

\end{quote}

\vspace*{\fill}
\begin{flushleft}
December 2013

\end{flushleft}
\end{titlepage}

\setcounter{footnote}{1}
\renewcommand{\thefootnote}{\fnsymbol{footnote}}

\section{Introduction}

The investigation of the origin of the electroweak symmetry breaking
is one of the main goals of the physics program at the Large Hadron Collider (LHC). 
The detailed study of the scalar resonance recently
discovered by the ATLAS and CMS experiments \cite{Aad:2012tfa,Chatrchyan:2012ufa} 
could lift the veil on the fundamental mechanism that gives mass to the known elementary particles. 

One of the important production mechanisms of a light Higgs boson \cite{Higgs:1964ia,Englert:1964et} at hadron colliders
is the associated production with a vector boson $V=W^\pm,Z$ (also known as the Higgs-strahlung process).
The vector boson provides a clean experimental signature, 
due to the presence of a high-$p_T$ lepton(s) and/or large missing transverse energy,
and allows us to tag the $H\to b \bar b$ decay, which is characterized by a large branching fraction.
This channel offers the opportunity to separately study the Higgs couplings to $W$ and $Z$ bosons.
The \VH\ production
was the main search channel for a light Higgs boson at the Tevatron,
and lead to the observation of an excess of events \cite{Aaltonen:2012qt}
compatible with the scalar resonance observed at the LHC.

At the LHC the associated \VH\ production was considered less promising, due to
the large backgrounds.
This situation can be substantially
improved by restricting the analysis to the so called
{\em boosted} region, where the vector boson and/or the $b \bar b$ pair have a large transverse momentum,
and possibly applying an extra light-jet veto~\cite{Butterworth:2008iy}.
This search strategy, however, significantly reduces the number of signal events,
and its potential will be fully exploited only when
the centre-of-mass energy $\sqrt{s}$ will reach 13 (14) TeV.
At present, with the full LHC data set at $\sqrt{s}=7$ and 8 TeV essentially analysed,
ATLAS \cite{ATLAS:2013-079} sees no signal in this channel, with a signal strength,
relative to that of the Standard Model (SM) Higgs boson, which is
$\mu=0.2\pm 0.5~{\rm (stat.)}\pm 0.4~{\rm (syst.)}$.
CMS \cite{Chatrchyan:2013zna}
sees a (small) excess of events above the expected background with a local significance of $2.1\sigma$,
consistent with the expectation from the production of the SM Higgs boson.
The signal strength corresponding to this excess is $\mu=1.0\pm 0.5$.

The actual experimental analyses are based on complicated selection cuts and
it is thus important to count on an accurate modelling of QCD radiation.
In order to obtain good control of the efficiency of the selection cuts,
and to assess whether the Monte Carlo tools
correctly describe the relevant distributions,
(fully) differential computations including the available radiative corrections are necessary.

The status of theoretical predictions for \VH\ production goes as follows.
The NNLO QCD corrections for the \VH\ inclusive cross section were computed in \cite{Brein:2003wg},
where all the Drell--Yan-like \cite{Hamberg:1990np} contributions (plus the gluon induced heavy-quark mediated corrections for the \ZH\ case)
were included. The quark induced heavy-quark mediated corrections for the \VH\ 
inclusive cross section were computed in \cite{Brein:2011vx} and found to be at the 1-3\% level at the LHC.
Soft-gluon effects to \VH\ production have been considered in Ref.~\cite{Dawson:2012gs}.
A fully differential computation of NNLO QCD corrections for \WH\ production was presented
in Ref.~\cite{Ferrera:2011bk}, while in Ref.~\cite{Banfi:2012jh} the NLO QCD corrections for both \WH\ production and
$H\to b\bar b$ decay were combined.
The NLO electroweak corrections for  \WH\ production have been computed \cite{Ciccolini:2003jy}
and implemented in the fully exclusive numerical code {\sc HAWK}.
The computation of the
fully differential $H\to b{\bar b}$ decay rate in NNLO QCD
has been reported in Ref.~\cite{Anastasiou:2011qx}.
The inclusive $H\to b{\bar b}$ decay rate is
known up to ${\cal O}(\as^4)$ \cite{Baikov:2005rw}.

As far as Monte Carlo implementations are concerned,
NLO corrections to $V\!H$ production have been matched to the parton shower within
the \MC@NLO \cite{Frixione:2002ik} framework  in Ref.~\cite{LatundeDada:2009rr} and within the
\POWHEG\ \cite{Frixione:2007vw} framework in Ref.~\cite{Hamilton:2009za}.
Recently, an NLO simulation matched to the parton shower for $V\!H+1$ jet has been presented in Ref.~\cite{Luisoni:2013cuh}, and merged by using the method of Ref.~\cite{Hamilton:2012rf}, with the corresponding $V\!H+0$ jet simulation.
At present, the ATLAS analysis \cite{ATLAS:2013-079} is based on a Monte Carlo signal sample generated with PYTHIA8 \cite{Sjostrand:2007gs}, whereas the CMS analysis \cite{Chatrchyan:2013zna} uses
\POWHEG\ interfaced with \HERWIG++\ \cite{Bahr:2008pv}\footnote{We note that \HERWIG++\ includes the possibility to account for NLO corrections in both $V\!H$ production and $H\to b{\bar b}$ decay \cite{Richardson:2012bn}, but
the Monte Carlo sample used by CMS is generated with a LO $H\to b{\bar b}$ decay.}.

As it was shown in Ref.~\cite{Ferrera:2011bk}, even if the effect of higher orders QCD radiative corrections 
can be relatively modest on the inclusive cross section 
 \cite{Brein:2003wg}, its impact on the accepted cross section and the relevant kinematical distributions
can be quite significant, in particular when severe selection cuts are applied, as
in the boosted \VH\ analysis.
The calculation of Ref.~\cite{Ferrera:2011bk} considered QCD corrections only to the production process $pp\to W\!H$, by neglecting QCD radiative effect in the \Hbb\ decay.
Fully inclusive QCD effects in the \Hbb\ decay were taken into account by normalizing the \Hbb\ branching fraction to the result of Ref.~\cite{Dittmaier:2011ti}.
This should be a good approximation
if one considers observables that are sufficiently inclusive over the extra radiation from the $b{\bar b}$ pair.
The study of Ref.~\cite{Banfi:2012jh}, however, casts some doubts on this approximation, by showing that QCD effects
from the decay can be relatively important,
especially with the selection cuts used by the LHC experiments at $\sqrt{s}=8$ TeV.

In this paper we extend and update the analysis presented in Ref.~\cite{Ferrera:2011bk} in two respects.
As a first step towards a complete NNLO calculation of QCD corrections for $pp\to W\!H\to l\nu b{\bar b}$,
we supplement the NNLO calculation of Ref.~\cite{Ferrera:2011bk} with QCD corrections to the \Hbb\ decay up to NLO.
As mentioned above, one important point is to understand the extent to which the QCD radiative effects are captured by the Monte Carlo generators used in the analysis. We thus compare our fixed order results with those obtained
with the \MC@NLO event generator, which includes radiation from the $b{\bar b}$ pair through the parton shower.
Our analysis is performed both at $\sqrt{s}=8$ and 14 TeV, by using the selection cuts typically applied
by the ATLAS and CMS collaborations.

This paper is organized as follows. In Sect.~\ref{sec:compu} we describe our calculation.
In Sect.~\ref{sec:lhc8} we present our results at the LHC with $\sqrt{s}=8$ TeV, and in Sect.~\ref{sec:lhc14}
we consider the case of the LHC with $\sqrt{s}=14$ TeV.
In Sect.~\ref{sec:summa} we summarize our results.

\section{Computation}
\label{sec:compu}

In this Section we introduce the theoretical framework adopted in our calculation. We consider the inclusive hard scattering process
\begin{equation}
\label{process}
pp\to W\!H+X\to W b{\bar b}+X\, ,
\end{equation}
where the Higgs boson $H$, which subsequently decays into a $b{\bar b}$ pair,
is produced together with a $W$ boson\footnote{The leptonic decay of the $W$ boson (including spin correlations) does not lead to complications and is understood in this Section to simplify the notation.}.
Our goal is to construct
the most precise predictions for the distributions
that are sensitive to selection cuts and vetoes on the jet activity
in both the production and decay stages of the Higgs boson.

The production differential cross section for the process (\ref{process}) can be written as:
\begin{equation}
d\sigma_{pp \rightarrow W\!H+X}=
d\sigma^{(0)}_{pp \rightarrow W\!H+X}+
d\sigma^{(1)}_{pp \rightarrow W\!H+X}+
d\sigma^{(2)}_{pp \rightarrow W\!H+X}+{\cal O}(\as^3)\, ,
\end{equation}
where $d\sigma^{(0)}$ is the LO contribution, and $d\sigma^{(1)}$ and $d\sigma^{(2)}$
the NLO and NNLO correction, respectively.
Analogously, the $H\to b{\bar b}$ differential decay rate is
\begin{equation}
d\Gamma_{H \rightarrow b \bar{b}}=
d\Gamma^{(0)}_{H \rightarrow b \bar{b}}+
d\Gamma^{(1)}_{H \rightarrow b \bar{b}}+d\Gamma^{(2)}_{H \rightarrow b \bar{b}}+{\cal O}(\as^3)\, .
\end{equation}
By using the narrow width approximation, the differential cross section for (\ref{process}) can
be written as
\begin{equation}
\label{master}
d\sigma_{pp \rightarrow W\!H+X\to W b{\bar b}+X} =
\left[
\sum_{k=0}^\infty d\sigma^{(k)}_{pp \rightarrow W\!H+X}\right] \times
\left[\frac{\sum_{k=0}^\infty d\Gamma^{(k)}_{H \rightarrow b \bar{b}}}
{\sum_{k=0}^\infty \Gamma^{(k)}_{H \rightarrow b \bar{b}}}\right]
\times Br(H \to b \bar{b})\, .
\end{equation}
Through Eq.~(\ref{master}) we can exploit the precise prediction of the Higgs
boson branching ratio into $b$ quarks $Br(H\to b{\bar b})$,
reported in \cite{Dittmaier:2011ti}, by which we 
normalize the contributions to the
differential decay rate of the Higgs boson.
We can consider various approximations of Eq.~(\ref{master}).
We first consider NLO corrections to the production process and ignore
QCD corrections to the decay, by defining
\begin{equation}
\label{eqnloprod}
d\sigma_{pp \rightarrow W\!H+X\to W b{\bar b}+X}^{\rm NLO(prod)+LO(dec)}=
\left[d\sigma^{(0)}_{pp \rightarrow W\!H+X}
+d\sigma^{(1)}_{pp \rightarrow W\!H+X}\right]\times 
d\Gamma^{(0)}_{H \rightarrow b \bar{b}}/\Gamma^{(0)}_{H \rightarrow b \bar{b}}
\times Br(H \to b \bar{b})\, .
\end{equation}
By including NLO corrections to the \Hbb\ decay we define
\begin{equation}
\label{eqnloproddec}
d\sigma_{pp \rightarrow W\!H+X\to W b{\bar b}+X}^{\rm NLO(prod)+NLO(dec)}=
\left[
d\sigma^{(0)}_{pp \rightarrow W\!H} \times
\frac{d\Gamma^{(0)}_{H \rightarrow b \bar{b}}+d\Gamma^{(1)}_{H \rightarrow b \bar{b}}}
{\Gamma^{(0)}_{H \rightarrow b \bar{b}}+\Gamma^{(1)}_{H \rightarrow b \bar{b}}}  +
d\sigma^{(1)}_{pp \rightarrow W\!H+X}
\times
\frac{d\Gamma^{(0)}_{H \rightarrow b \bar{b}}}
{\Gamma^{(0)}_{H \rightarrow b \bar{b}}}
\right] \times Br(H \rightarrow b \bar{b})\, ,
\end{equation}
which represents the complete NLO calculation considered in Ref.~\cite{Banfi:2012jh}.
We point out here that at the first order in $\as$ the factorization between
production and decay is indeed exact because of colour conservation.
In other words the interference of QCD
radiation in Higgs boson production and decay stages vanishes at
this order. This property does not hold beyond ${\cal O}(\as)$.

As a first step towards a complete NNLO calculation
we consider the following approximation of Eq.~(\ref{master})
\begin{align}
\label{eqnnlo}
d\sigma_{pp \rightarrow W\!H+X\to l\nu b{\bar b}+X}^{\rm NNLO(prod)+NLO(dec)} =&
\left[
d\sigma^{(0)}_{pp \rightarrow W\!H} \times
\frac{d\Gamma^{(0)}_{H \rightarrow b \bar{b}}+d\Gamma^{(1)}_{H \rightarrow b \bar{b}}}
{\Gamma^{(0)}_{H \rightarrow b \bar{b}}+\Gamma^{(1)}_{H \rightarrow b \bar{b}}} \right. \nonumber \\
&+
\left.
\left(
d\sigma^{(1)}_{pp \rightarrow W\!H+X}+
d\sigma^{(2)}_{pp \rightarrow W\!H+X}
\right) \times
\frac{d\Gamma^{(0)}_{H \rightarrow b \bar{b}}}
{\Gamma^{(0)}_{H \rightarrow b \bar{b}}}
\right] \times Br(H \rightarrow b \bar{b})\, .
\end{align}
In Eq.~(\ref{eqnnlo}) we include QCD corrections to the production stage up to NNLO,
and the Higgs decay is treated up to NLO.
Although this is not a fully consistent approximation,
since it neglects some ${\cal O}(\as^2)$ contributions in Eq.~(\ref{master}),
we believe it captures the relevant radiative effects (see discussion below).

The NNLO computation for the production process \cite{Ferrera:2011bk}
is performed using the subtraction
method proposed in~\cite{Catani:2007vq}.
This method allows us to compute up to NNLO contributions in QCD for the whole
class of hadronic collisions producing a colourless final state at LO and it
has been successfully applied to the computation of
NNLO corrections to several hadronic processes \cite{Catani:2007vq,Catani:2009sm,Catani:2011qz,Grazzini:2013bna}.

The $H\to b{\bar b}$ decay at NLO is computed by using the dipole subtraction method \cite{Catani:1996jh,Catani:1996vz,Catani:2002hc} and is included in a fully differential numerical code both for massless and massive $b$ quarks.
We point out that
in the on shell scheme
the heavy-quark mass dependence leads to large logarithmic terms of the form $\ln m_H/m_b$, which render the whole $H\to b{\bar b}$ decay rate infrared unsafe.
To correctly recover the $m_b\to 0$ limit these logarithmic terms must be absorbed into the running $Hb{\bar b}$ Yukawa coupling \cite{Braaten:1980yq,Drees:1990dq}.
With this treatment the massless and massive computations produce in practice almost identical results\footnote{We note that the authors of Ref.~\cite{Banfi:2012jh} do not absorb these large logarithmic terms into the $Hb{\bar b}$ coupling and this leads to differences in the quantitative results of the two NLO computations.}, and in the next Section we thus limit ourselves to consider the massless case.

\section{Numerical results at $\sqrt{s}=8$ TeV}
\label{sec:lhc8}

In this Section we present numerical results at the LHC,
in the case $\sqrt{s}=8$~TeV.
We thus consider \WH\ production in $pp$ collisions
followed by the $W \rightarrow l\,\nu_l$ and $H\to b{\bar b}$ decays.
We first focus on the $p_T$ spectrum of the Higgs candidate,
whose knowledge is particularly important in the experimental analysis,
and then we present our results for the corresponding
invariant mass distribution.
We use the so called $G_\mu$ scheme for the electroweak couplings,
where the input parameters are $G_F$, $m_Z$, $m_W$. In particular we 
use the following values:
$G_F = 1.16637\times 10^{-5}$~GeV$^{-2}$,
$m_Z = 91.1876$~GeV, $m_W = 80.399$~GeV and $\Gamma_W=2.085$~GeV.
The mass of the SM Higgs boson is set to $m_H=125$~GeV
and the width to $\Gamma_H=4.070$ MeV~\cite{Dittmaier:2011ti}.
The \WH\ production cross section is computed up to NNLO by using the calculation of Ref.~\cite{Ferrera:2011bk},
including the leptonic decay of the $W$ boson,
in both the $W^+\to l^+\nu_l$ and $W^-\to l^-{\bar \nu}_l$ channels.
We compute the $H\to b{\bar b}$ decay up to NLO QCD
and we normalize the $Hb{\bar b}$ Yukawa coupling such that the value of the branching ratio is
$Br(H\to b{\bar b})=0.578$ \cite{Dittmaier:2011ti}.
The fixed order results are compared with the results obtained with the \MC@NLO 4.09 event generator \cite{Frixione:2002ik}, normalized to the same $H\to b{\bar b}$ branching ratio, and without underlying event.

As for the parton distribution functions (PDFs) we use the NNPDF2.3 PDF sets \cite{Ball:2012cx}, with
densities and $\as$ evaluated at each corresponding order
(i.e., we use $(n+1)$-loop $\as$ at N$^n$LO, with $n=0,1,2$) and with $\as(m_Z)=0.118$.
In the fixed order calculations
the central values of the renormalization and factorization scales are fixed to the value 
$\mu_R=\mu_F=m_W+m_H$ while the central value of the renormalization scale for the 
$H\to b{\bar b}$ coupling is set to the value $\mu_r=m_H$.
In the \MC@NLO simulation the central scale is the default scale, the
transverse mass of the $W\!H$ system.
Jets are reconstructed with the anti-$k_T$ algorithm 
with $R=0.4$ \cite{Cacciari:2008gp} and with a transverse momentum $p_T^j>20$~GeV. 
In order to simulate the experimental analysis for the Higgs search in this channel, 
we require exactly two ($R$) separated $b$-jets each with $p_T^b > 30$~GeV and $|\eta_b|<2.5$. 
In the fixed-order calculation a jet is considered a $b$-jet if it contains at least one $b$-quark. In the \MC@NLO simulation this is achieved by requiring that, after hadronization, the jet contains at least one $B$-hadron.

\begin{figure}[th]
\begin{center}
\begin{tabular}{cc}
\includegraphics[width=.50\textwidth]{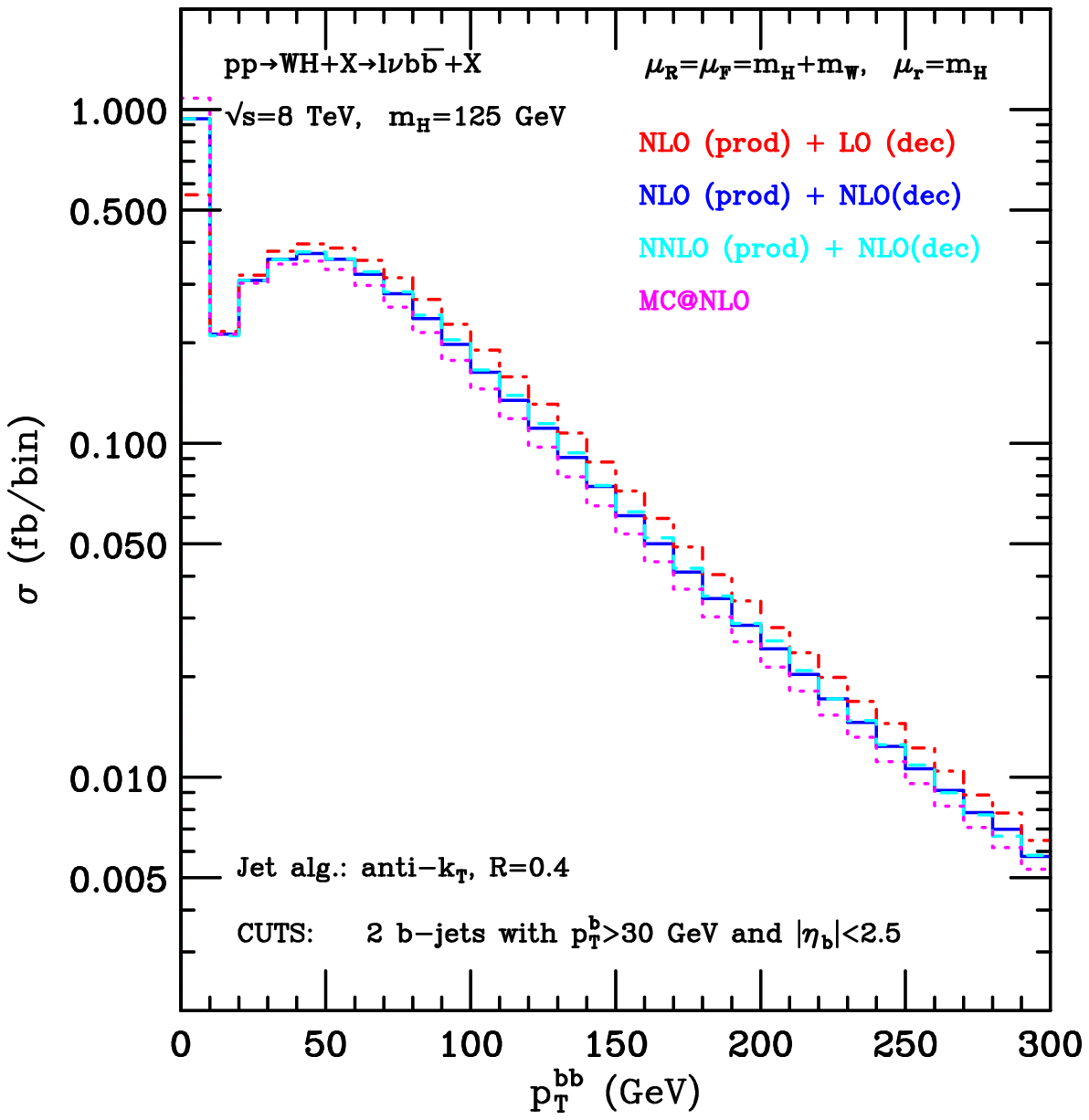}
\includegraphics[width=.45\textwidth]{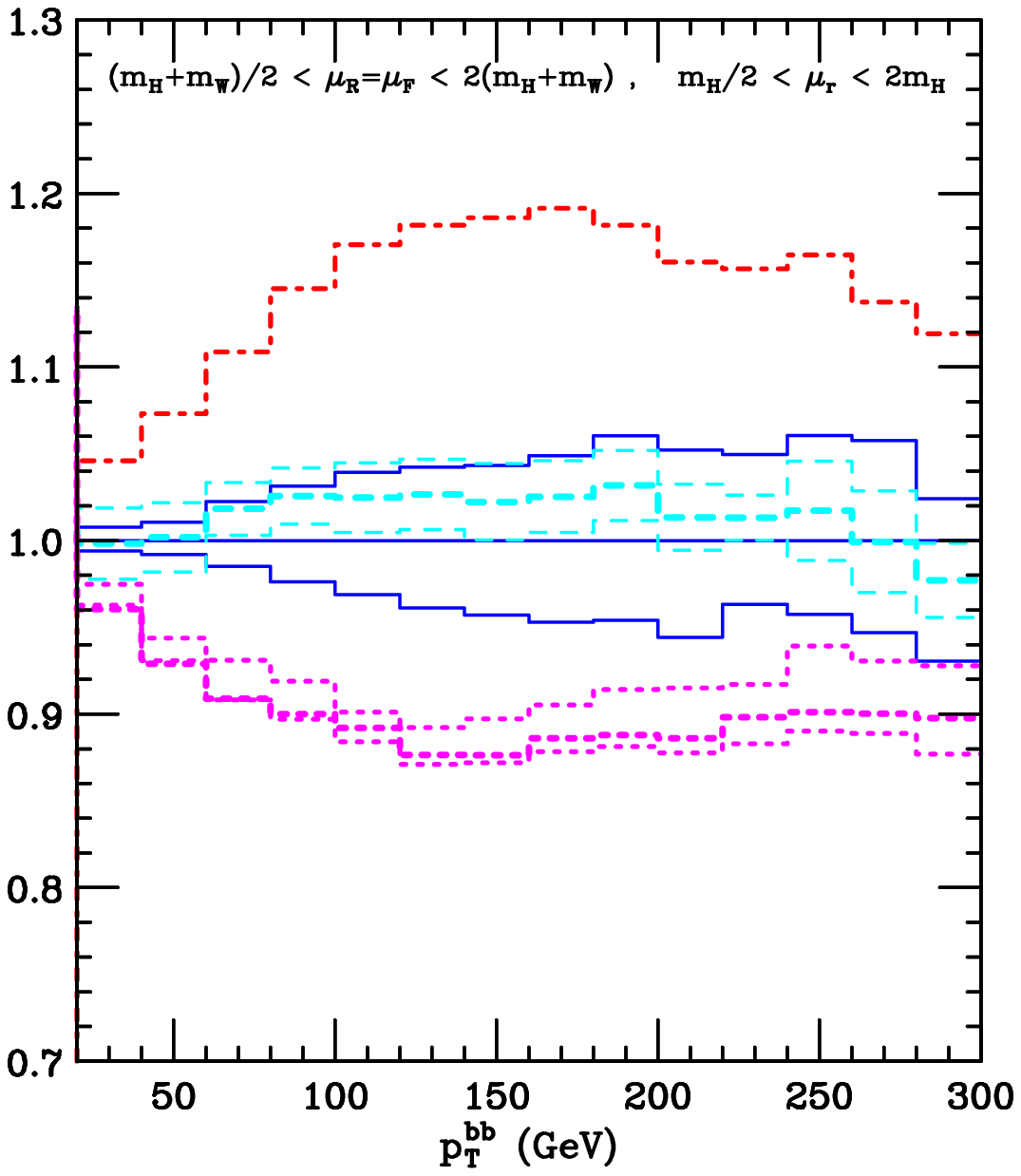}
\end{tabular}
\end{center}
\caption{\label{fig:nocuts}
{\em Left panel: Transverse-momentum distribution of the $b$-jets pair computed at NLO with LO decay (red dot-dashes), NLO with NLO decay (blue solid), NNLO with NLO decay (cyan dashes) and with \MC@NLO (magenta dots). Right panel: The same distributions normalized to the full NLO result. The NLO, NNLO and \MC@NLO uncertainty bands are also shown. No cuts except the $b$-jet selection are applied.}}
\end{figure}

We start the presentation of our results
by considering the inclusive \WH\ selection of the $b$-jet pair.
Note that thanks to
Eq.~(\ref{master}) and to the normalization of the $H\to b{\bar b}$ coupling, 
 the prediction for the total cross-section is insensitive to the 
higher-order corrections to the $H\to b{\bar b}$ decay for a {\it completely} inclusive quantity: this is a valuable check
of the implementation of the NLO corrections to the Higgs
boson decay (still we can observe differences in the shape of distributions).
In this inclusive case the only effective selection cuts are the minimum value of the transverse momentum used in the 
jet definition  and the cuts which define the separated
$b$-jets. 
In Fig.~\ref{fig:nocuts} (left panel) we show the QCD predictions at NLO (with and without NLO corrections
to the $H\to b{\bar b}$ decay, see Eqs.~(\ref{eqnloprod}), (\ref{eqnloproddec})), at NNLO (see Eq.~(\ref{eqnnlo})) and  from \MC@NLO, for the 
transverse-momentum distribution of the $b$-jets pair $p_T^{b\bar b}=|\vec{p}_T^{\,\,b}+\vec{p}_T^{\,\,\bar b}|$.
In the $p_T^{b\bar b}=0$ bin we collect the events which do not fulfil the selection cuts.
Here and in the following we take the complete NLO result (see Eq.~(\ref{eqnloproddec})) as reference theoretical prediction and
in Fig.~\ref{fig:nocuts} (right panel) we plot the NLO, NNLO and \MC@NLO $p_T$ distributions normalized to the NLO result, with their scale uncertainty band, which is obtained as follows. In the fixed order calculations we vary $\mu_F=\mu_R$ between $(m_H+m_W)/2$ and $2(m_H+m_W)$ and, simultaneously, we vary the decay scale $\mu_r$ between $m_H/2$ and $2m_H$. In \MC@NLO $\mu_F=\mu_R$ is varied by a factor of two around the
central scale, the transverse mass of the $W\!H$ system.

By comparing the different spectra in Fig.~\ref{fig:nocuts}
we see that the hardest
is the NLO one (with LO $H\to b{\bar b}$ decay), with a selection efficiency of the $b$-jet pair of $88\%$.
If we consider the full NLO corrections,
the spectrum becomes softer and the efficiency decreases to $79\%$.
This is not unexpected since, generally speaking, hard real emissions from the $b{\bar b}$ pair reduce
the $p_T^{b\bar b}$ of the 
event and increase the probability that the $b$-quark radiating a hard gluon could fail the $p_T^b> 30$~GeV 
threshold. This situation does not change significantly if we further consider the NNLO corrections for 
the production: we observe only a slight increase of the accepted cross section, at the $1\%$
level.
The effect of scale variations at NLO (NNLO) is of the order of about $\pm 2\%$ ($\pm 1\%$) on the accepted cross section, 
but it increases at high $p_T$, where it can be of ${\cal O}(\pm 5\%)$ (${\cal O}(\pm 3\%)$).
The \MC@NLO prediction, besides the NLO plus parton shower effects for the production, includes radiation from the $b{\bar b}$ pair due to the parton shower.
In this case, we observe 
that even if the matrix elements for the  $H\to b{\bar b}$ decay have a LO accuracy, the effect of the shower 
is qualitatively similar to (but quantitative larger than)
the NLO corrections to the decay: the spectrum is softer
and the efficiency reaches the $75\%$ level. The physical picture is the one discussed before:
parton emissions from the $b{\bar b}$ pair reduce the $p_T^{b\bar b}$ of the event and decrease the efficiency.

\begin{figure}[th]
\begin{center}
\begin{tabular}{cc}
\includegraphics[width=.50\textwidth]{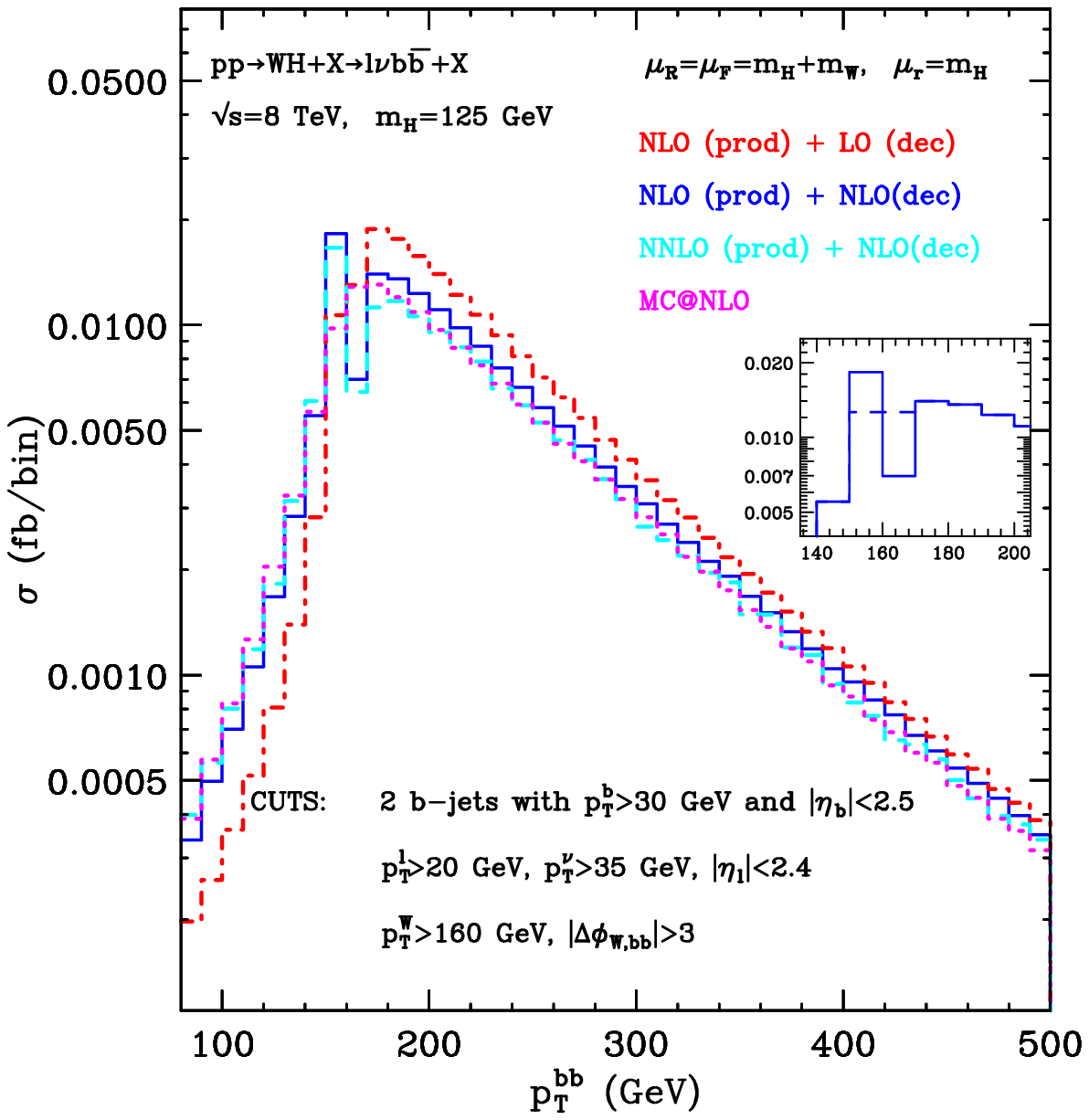}
\includegraphics[width=.45\textwidth]{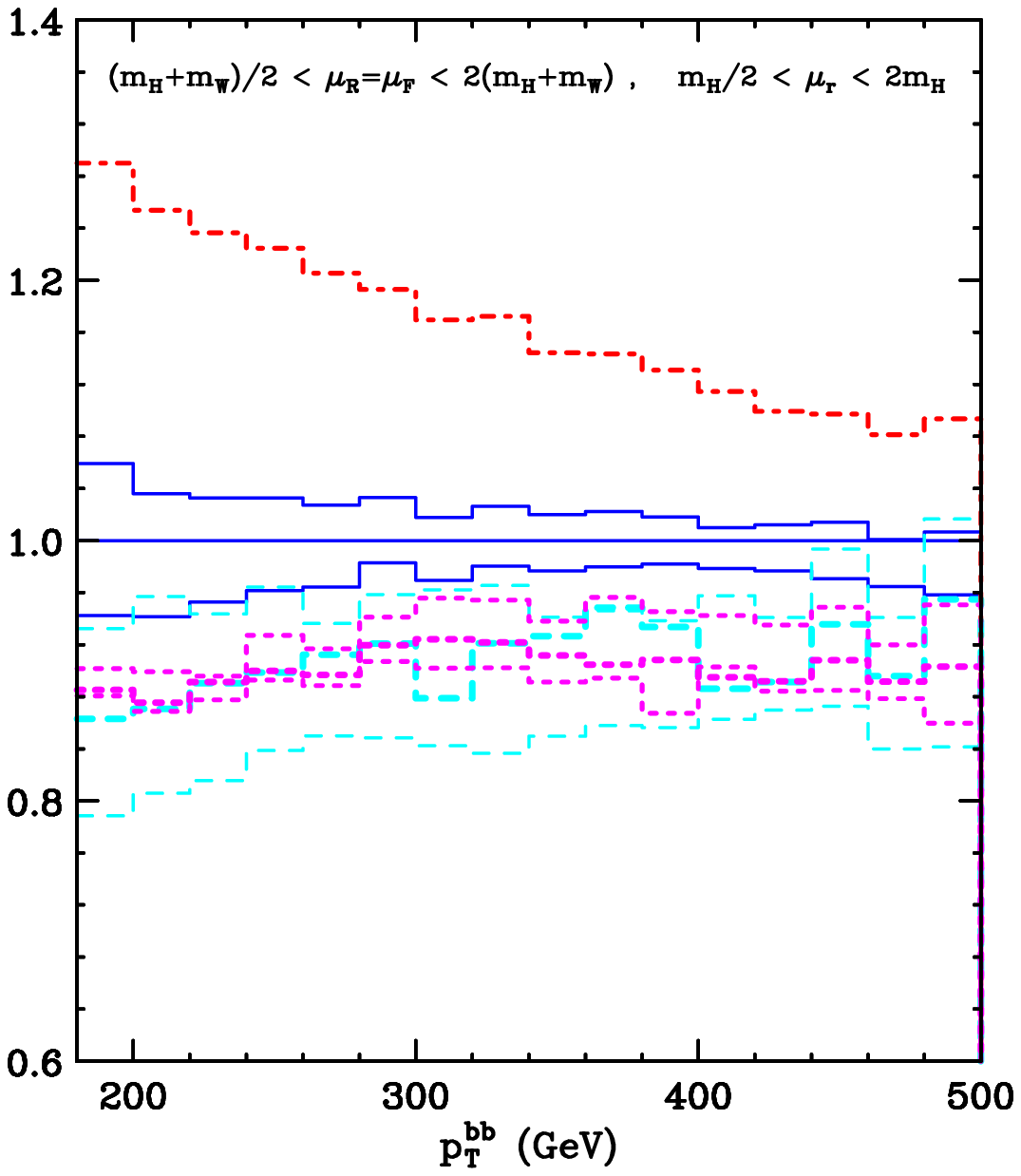}
\end{tabular}
\end{center}
\caption{\label{fig:cuts}
{\em As in Fig.~\ref{fig:nocuts} but when selection cuts are applied. The inset plot shows the region around $p_T^{bb}\sim 160$ GeV.}}
\end{figure}

We now proceed to consider a more realistic situation in which
we apply selection cuts similar to those used by ATLAS and CMS in their analysis.
At $\sqrt{s}=8$ TeV and with the integrated luminosity accumulated, it
is not really possible to perform a boosted analysis like that
proposed in Ref.~\cite{Butterworth:2008iy}.
The strategy of the Higgs boson search in this channel is thus to apply less stringent
selection cuts, which aim at having the Higgs and the $W$ boson at relatively large $p_T$, and almost back to back, to reduce the $t{\bar t}$ background.

In particular, we consider here the following cuts.
The charged lepton is required to have transverse momentum $p_T^l > 20$~GeV
and pseudorapidity $|\eta_l|< 2.4$;
the missing transverse momentum of the event is required to be 
$p_T^\nu>35$~GeV.
The $W$ boson must have a transverse momentum $p_T^W > 160$~GeV and
is required to be almost back-to-back with the Higgs candidate. To achieve
this condition the azimuthal separation of the $W$ boson with the $b \bar b$ pair must fulfil 
$|\Delta \phi_{W,bb}|>3$. The selection on $p_T^W$ is important to improve the signal-to-background ratio:
an analogous cut on the Higgs boson can be imposed by focusing
on the large $p_T$ region in the $p_T^{b\bar b}$ distribution.

In Fig.~\ref{fig:cuts} we study the $p_T^{b\bar b}$ distribution of the Higgs candidate.
As above we consider QCD predictions at NLO (with and without corrections
to the $H\to b{\bar b}$ decay), at NNLO (with NLO decay) and from \MC@NLO. The corresponding
cross sections and scale uncertainties are reported in the first row of Table~\ref{table1}.
As in Fig.~\ref{fig:nocuts}, in the right panel of Fig.~\ref{fig:cuts}
we plot the $p_T$ spectra normalized to the full NLO result. 

As in the inclusive case the hardest spectrum is the NLO one (with LO $H\to b{\bar b}$ decay), 
with an accepted cross section which is only $4\%$ with respect to the inclusive one (the bulk of the reduction
is due to the tight cut on $p_T^W$). When including
the NLO corrections to the $H\to b \bar b$ decay the spectrum becomes softer and the accepted cross section
is further reduced by $12\%$.

We observe from Fig.~\ref{fig:cuts} that the inclusion of the NLO corrections
produces  
instabilities around the region where $p_T^{b\bar b}=160$~GeV.
The origin of such instabilities is 
of Sudakov type~\cite{Catani:1997xc}:
at LO the $p_T^{W}>160$~GeV constraint imposes a kinematical boundary on the $p_T^{b\bar b}$ spectrum,
and perturbative contribution at higher orders produce
integrable logarithmic singularities around
such boundary.
The way to solve these perturbative instabilities is to perform an all-order resummation
of the soft-gluon contributions which renders the distribution smooth in the vicinity of 
the boundary. The effects of soft-gluon resummation can be mimicked by considering
a more inclusive observable i.e.\ by increasing the bin size of the 
distribution around the critical point. 
The effect of the smearing obtained in this way can be seen in the inset plot of Fig.~\ref{fig:cuts} (dashed line).

We also observe that the NLO corrections to the decay below the $p_T^{b\bar b}=160$~GeV  boundary
are particularly large  (reaching, for $p_T^{b\bar b}\sim 120$, the $100\%$ level with respect to the 
cross section with NLO corrections for the production only).
This is not unexpected, since in this region of transverse momenta,
the ${\cal O}(\as)$ correction to the Higgs boson decay contributes
as a leading order term.
Contrary to the inclusive case the NNLO corrections
for the production are not negligible:
the spectrum becomes softer and the accepted cross section
is further reduced by $9\%$. 

Comparing the fixed order predictions to the \MC@NLO result we observe 
that the effect of the shower is quantitative very similar 
to the effect of the NNLO corrections for the production plus NLO for the Higgs decay
(with the exclusion of the region around the LO kinematical boundary discussed before). 
Moreover we note that the \MC@NLO prediction
around the LO kinematical boundary has a smooth behaviour, without the instabilities  
of the fixed order case. 
This is because  the effective resummation of the Sudakov 
logarithms implemented 
in the shower algorithm  permits a more reliable description of the region around the boundary.

The NLO scale uncertainties are ${\cal O}(\pm 10\%)$ in the region $p_T\ltap 200$ GeV and then decrease
to the ${\cal O}(\pm 5\%)$ level or smaller for higher values of $p_T$.
From Fig.~\ref{fig:cuts} (right panel) we conclude that the inclusion of NLO corrections
to the Higgs decay is important
to obtain a reliable shape of the $p_T$ spectrum. The \MC@NLO prediction, on the other hand, even without the NLO corrections to the decay, describes the shape of the spectrum rather well. We also conclude that the NLO scale uncertainty is in this case
too small to be considered as a true uncertainty from missing higher order contributions, since both the NNLO and \MC@NLO results lie outside the NLO band. The NNLO uncertainty band is in turn larger than the NLO one,
being at the $\pm 7-8\%$ level, and marginally overlaps with the latter.
The NNLO and \MC@NLO results are perfectly compatible within the uncertainties.

\begin{figure}[th]
\begin{center}
\begin{tabular}{cc}
\includegraphics[width=.50\textwidth]{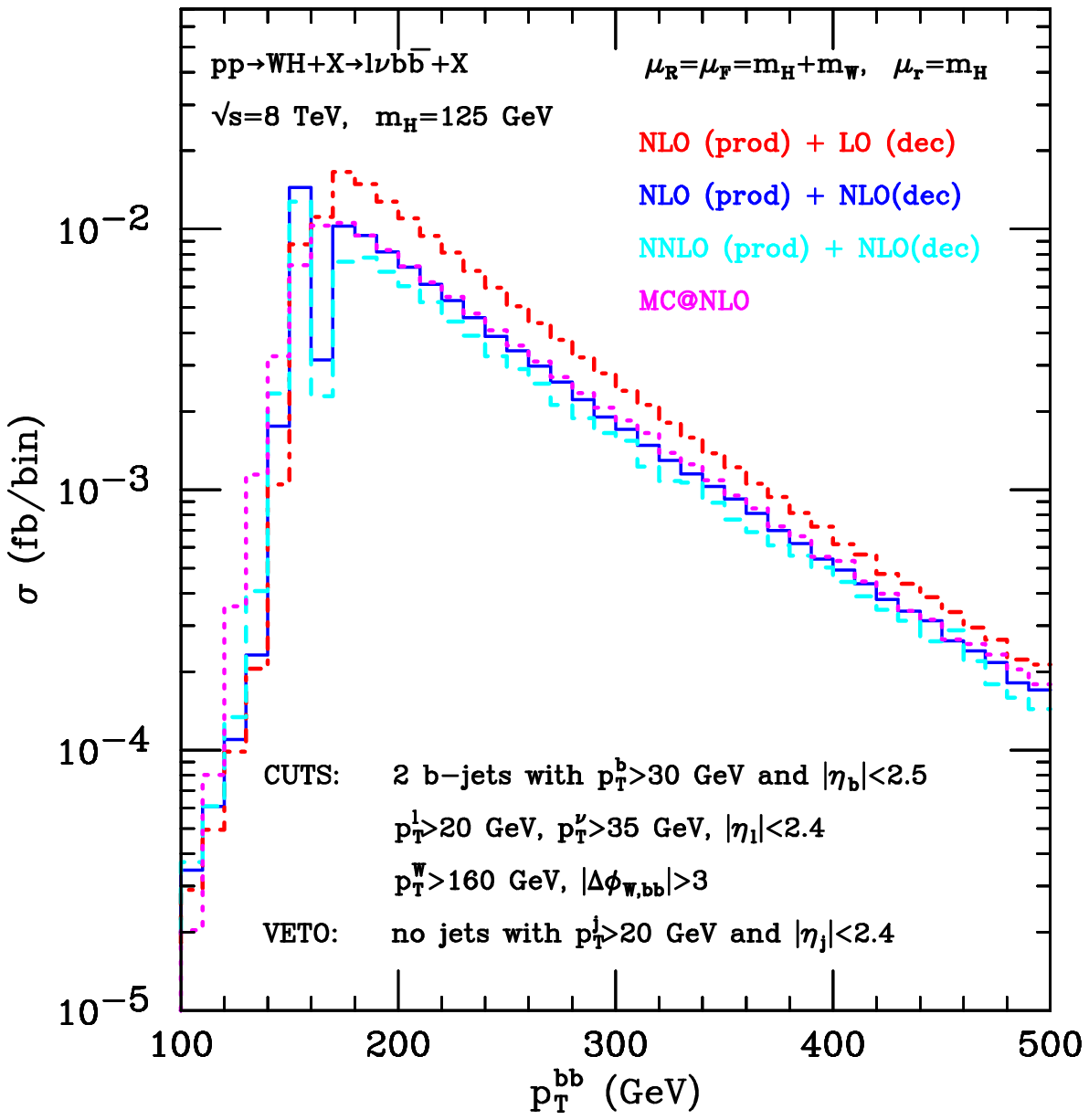}
\includegraphics[width=.46\textwidth]{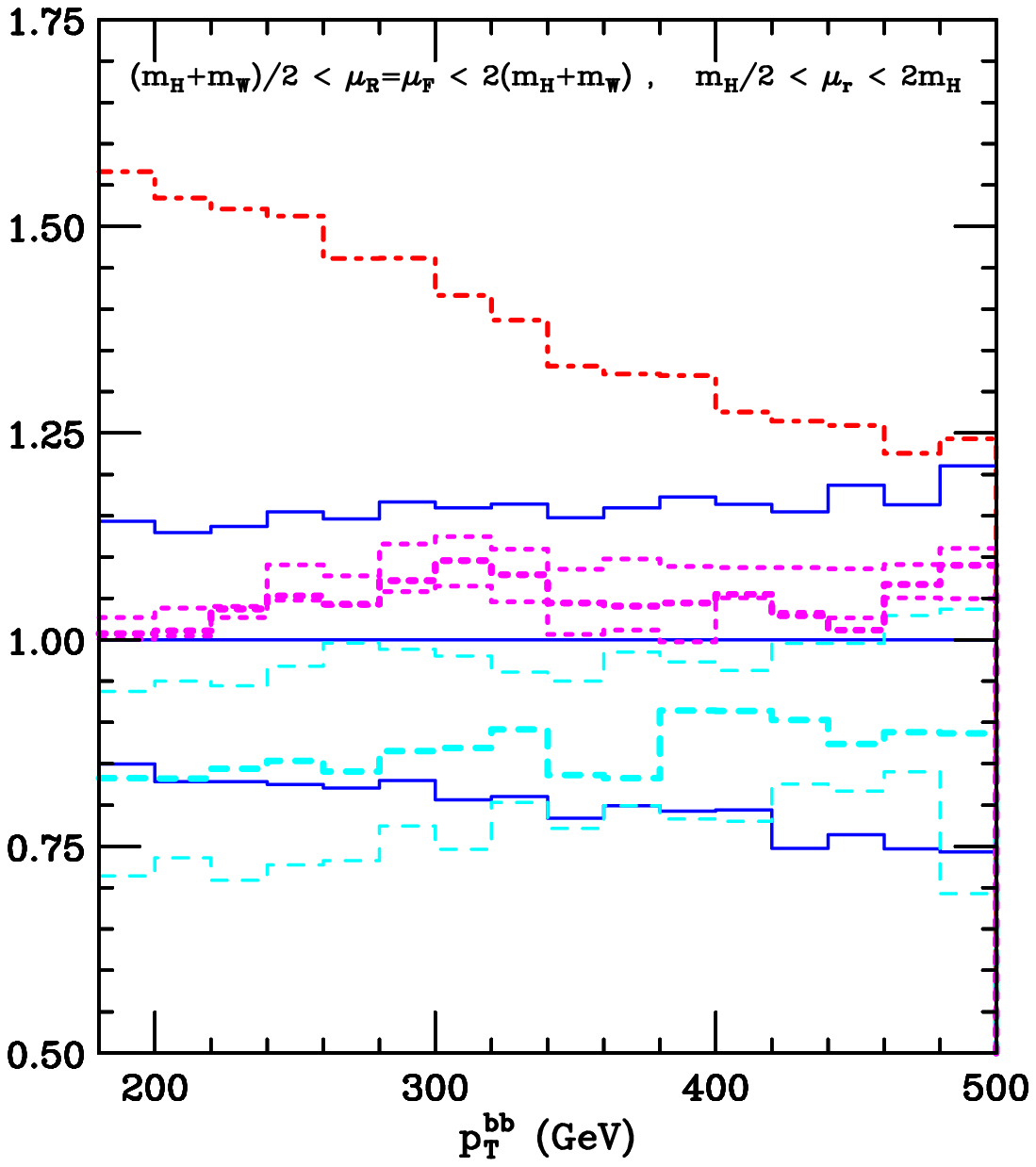}
\end{tabular}
\end{center}
\caption{\label{fig:vetocuts}
{\em As in Fig.~\ref{fig:cuts} but with an additional veto on light jets.}}
\end{figure}

To improve the background rejection, a veto on extra jet radiation is typically used in the analyses.
In Fig.~\ref{fig:vetocuts} we consider the case
in which, besides the cuts considered above, events with additional jets with $p_T^j > 20$~GeV
and pseudorapidity $|\eta_j|< 2.4$ are rejected.
The corresponding cross sections and scale uncertainties are reported in the second row of Table~\ref{table1}.
In order of increasing sensitivity, the effect of the jet veto is
to reduce the accepted cross section
by $25\%$ at NLO (production only),
by $33\%$ for \MC@NLO, by $41\%$ at full NLO accuracy
and by $44\%$ at the NNLO.
The reason of such sensitivity is the different content of radiative corrections which are present in
the calculations. Most sensitive to the jet veto is the NNLO distribution (with NLO Higgs decay) where up to two
hard emissions from the initial state and one hard emission from the final states are considered.
As a result the jet veto produces a different behaviour of the distributions with respect of the situation 
in  Fig.~\ref{fig:cuts}. In particular we observe that the full NLO result is 
very close to the \MC@NLO prediction while the inclusion of the NNLO corrections for the production further reduces the accepted cross section by $10\%$ (see Table~\ref{table1}). 

We add few comments on the stability of the perturbative results when a jet veto is applied \cite{Catani:2001cr}.
As is well known, when a generic system of high-mass $M$ is produced in hadronic collisions,
a veto on jets with $p_T>p_T^{\rm veto}$ leads to potential instabilities in the perturbative expansion,
since the cancellation between real and virtual contributions is unbalanced.
The typical scale of the accompanying QCD radiation is $\langle 1-z\rangle M$, where $1-z=1-M^2/{\hat s}$ is the average distance from the partonic threshold.
When this scale is larger than the 
jet veto scale $p_T^{\rm veto}$, the effect of the jet veto
is expected to be more sizeable.
The perturbative instabilities may originate from potentially
large logarithmic contributions of the form
$\ln (1-z) M/p_T^{\rm veto}$.
In our case (with $M=M_{W\!H}$ being the invariant mass of the $W\!H$ system)
the cuts already
select a phase space region in which the radiation recoiling against the \WH\ system
is relatively soft, and the additional reduction in the accepted cross section from the jet veto is limited.

As in Fig.~\ref{fig:nocuts} and \ref{fig:cuts}, in Fig.~\ref{fig:vetocuts} (right panel) we plot the $p_T$ spectra normalized to the reference NLO prediction, and
we study the scale uncertainties. The effect of NLO scale variations is definitely
larger than in Fig.~\ref{fig:nocuts} and \ref{fig:cuts}, being of the order of
${\cal O}(\pm 15-20\%)$ in the range considered.
We also see that the NNLO uncertainty is smaller than the NLO one,
being of ${\cal O}(\pm 10\%)$.
We point out that, contrary to what happens
without the jet veto (see Fig.~\ref{fig:cuts}), both the NNLO and \MC@NLO predictions lie within the NLO uncertainty band. This fact, together with the relatively mild impact of the jet veto on the accepted cross section, gives us confidence that the theoretical prediction is under good control.

\begin{table}[h]
\begin{center}
\begin{tabular}{|c|c|c|c|c|}
\hline
$\sigma$ (fb) & NLO (with LO dec.) & NLO (full) & NNLO (with NLO dec.) & \MC@NLO \\
\hline
\hline
w/o extra jet veto  & $1.96^{+1\%}_{-1\%}$ & $1.73^{+2\%}_{-3\%}$ &  $1.56^{+5\%}_{-5\%}$ & $1.58^{+2\%}_{-1\%}$ \\
\hline
w extra jet veto & $1.46^{+5\%}_{-8\%}$ & $1.02^{+14\%}_{-15\%}$ &  $0.87^{+11\%}_{-11\%}$ & $1.07^{+3\%}_{-1\%}$ \\
\hline
\end{tabular}
\end{center}
\caption{{\em Cross sections and their scale uncertainties
for $pp\to W\!H+X\to l\nu b \bar b+X$ at the LHC with $\sqrt{s}=8$ TeV. 
The applied cuts are described in the text.}}
\label{table1}
\end{table}

In Table~\ref{table1} we report the cross sections and scale uncertainties
obtained at the various orders, together with the \MC@NLO result.
The scale uncertainties
are obtained with the procedure discussed above.
We note that the \MC@NLO uncertainty turns out to be rather small.
In the case in which the jet veto is not applied (first row of Table~\ref{table1})
this is consistent with what we find at NLO (with LO decay).
When a jet veto is applied, the \MC@NLO uncertainty is still very small,
and smaller than the corresponding uncertainty of the NLO result,
thus suggesting that it could be underestimated.

\begin{figure}[th]
\begin{center}
\begin{tabular}{cc}
\includegraphics[width=.46\textwidth]{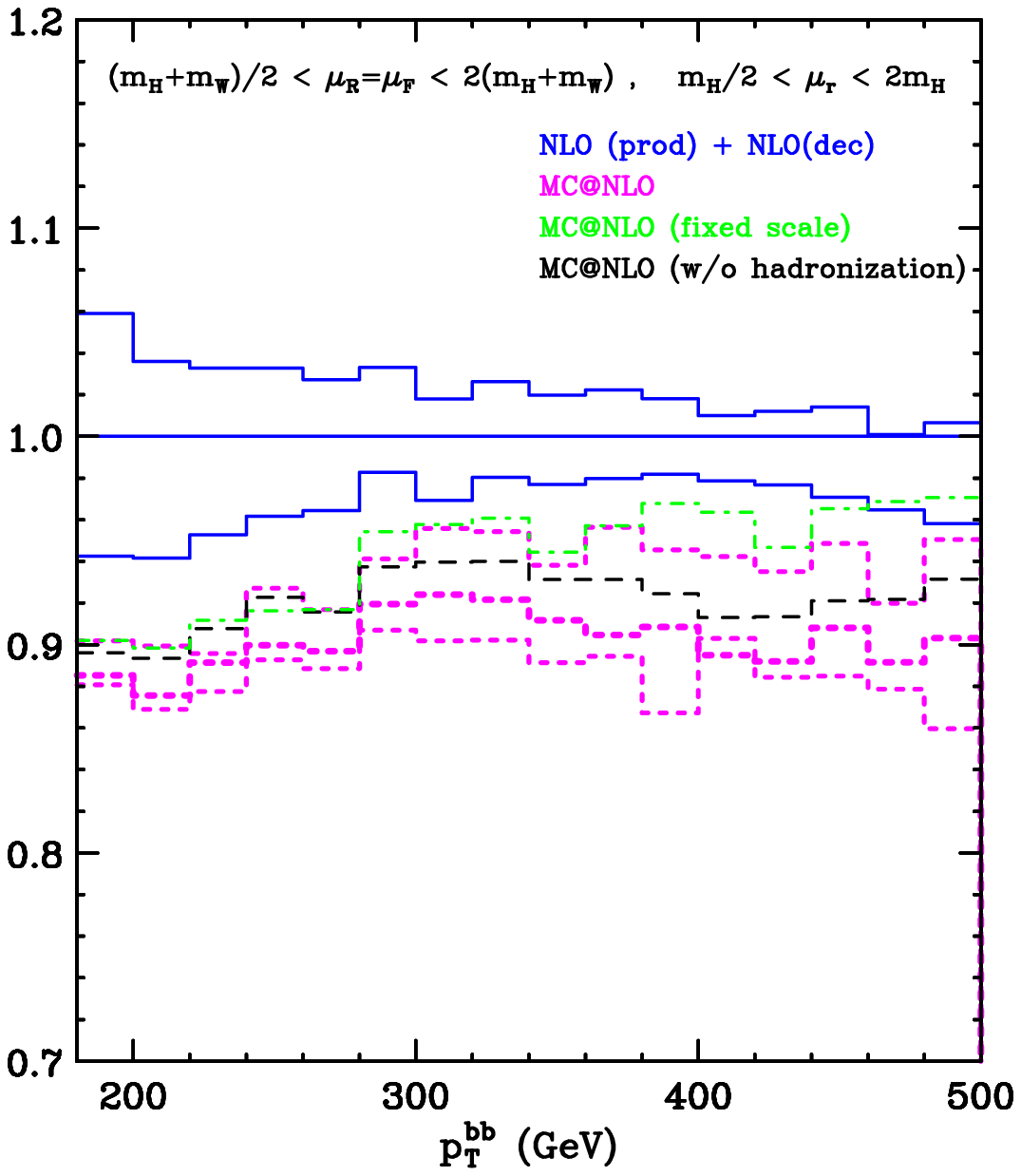}
\includegraphics[width=.46\textwidth]{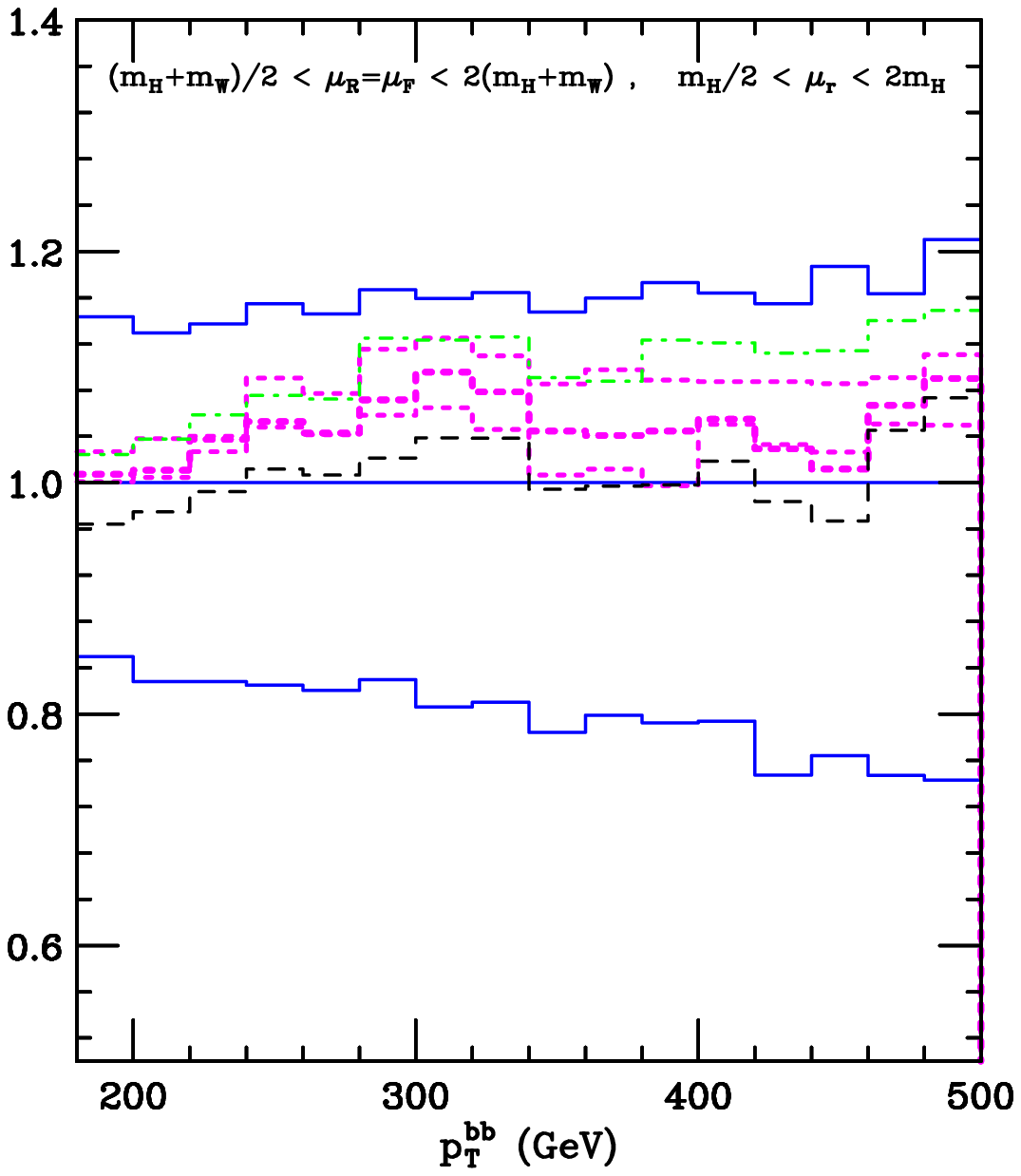}
\end{tabular}
\end{center}
\caption{\label{fig:nlomcatnlo}
{\em Comparison of NLO with NLO decay (blue solid), \MC@NLO with default scale (magenta dots), \MC@NLO with fixed scale (green dot-dashes), \MC@NLO without hadronization (black dashes). Left panel: without jet veto. Right panel: with jet veto.}}
\end{figure}

In the previous discussion we have compared results for the $p_T$ spectrum of the Higgs candidate obtained
at different perturbative orders with the result obtained with the \MC@NLO event generator, which uses, as default
scale, the transverse mass of the $W\!H$ system, and, besides the effect from the parton shower, includes hadronization. In order to disentangle these different effects in Fig.~\ref{fig:nlomcatnlo} we compare the default NLO and \MC@NLO results as in Figs.~\ref{fig:cuts} and \ref{fig:vetocuts}, with the \MC@NLO result obtained with $\mu_F=\mu_R=m_W+m_H$, and
with the \MC@NLO result without hadronization. The left panel corresponds to Fig.~\ref{fig:cuts} (no jet veto)
and the right panel corresponds to Fig.~\ref{fig:vetocuts} (with jet veto).
Comparing the \MC@NLO result with $\mu_F=\mu_R=m_W+m_H$ to the default one, we see that the former is generally consistent with the latter, and tends to lie at the upper edge of the default \MC@NLO band at high $p_T$.
This is consistent with the fact that the fixed scale leads to larger $\as$ and, as a consequence, larger
perturbative corrections at high $p_T$.
We see that the hadronization effects are relatively small, being at the $1-2\%$ level in the case in which
no jet veto is applied (left panel), and increase to the $5\%$ level when the jet veto is applied (right panel).
This is not unexpected: the Higgs $p_T$ spectrum is expected to be independent on hadronization if
the analysis is sufficiently inclusive. 
However there is a non trivial relation between the effects of the hadronization and the presence of a jet veto. 
Indeed, in Fig.~\ref{fig:nlomcatnlo} we observe that in the case in which the jet veto is applied (right panel) the contribution of the
hadronization has opposite sign with respect to the more inclusive case (left panel). This interplay between hadronization and jet veto
would require a dedicated study,  which, however, is beyond the scope of the present paper.

\begin{figure}[th]
\begin{center}
\begin{tabular}{cc}
\includegraphics[width=.48\textwidth]{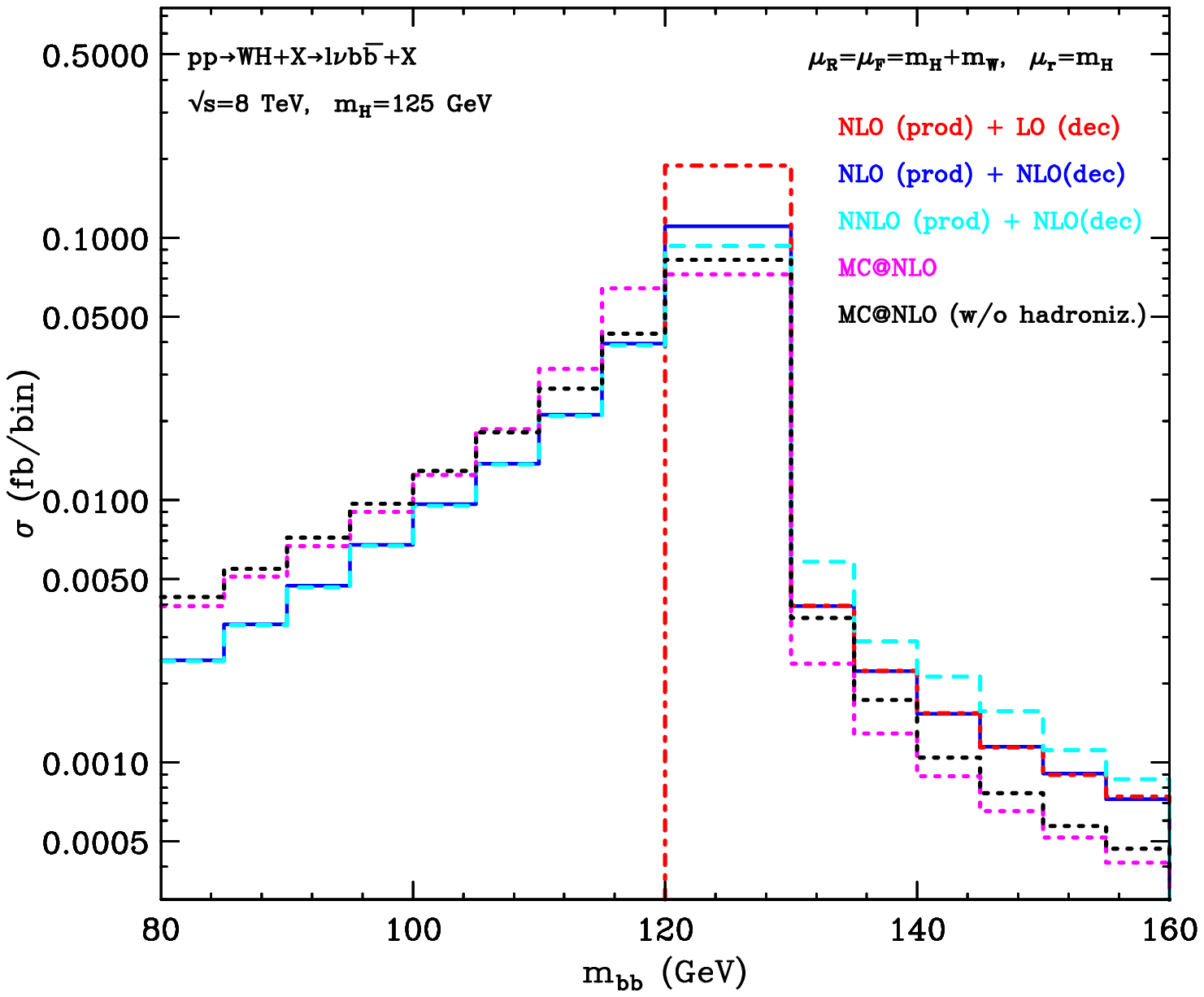}
\includegraphics[width=.48\textwidth]{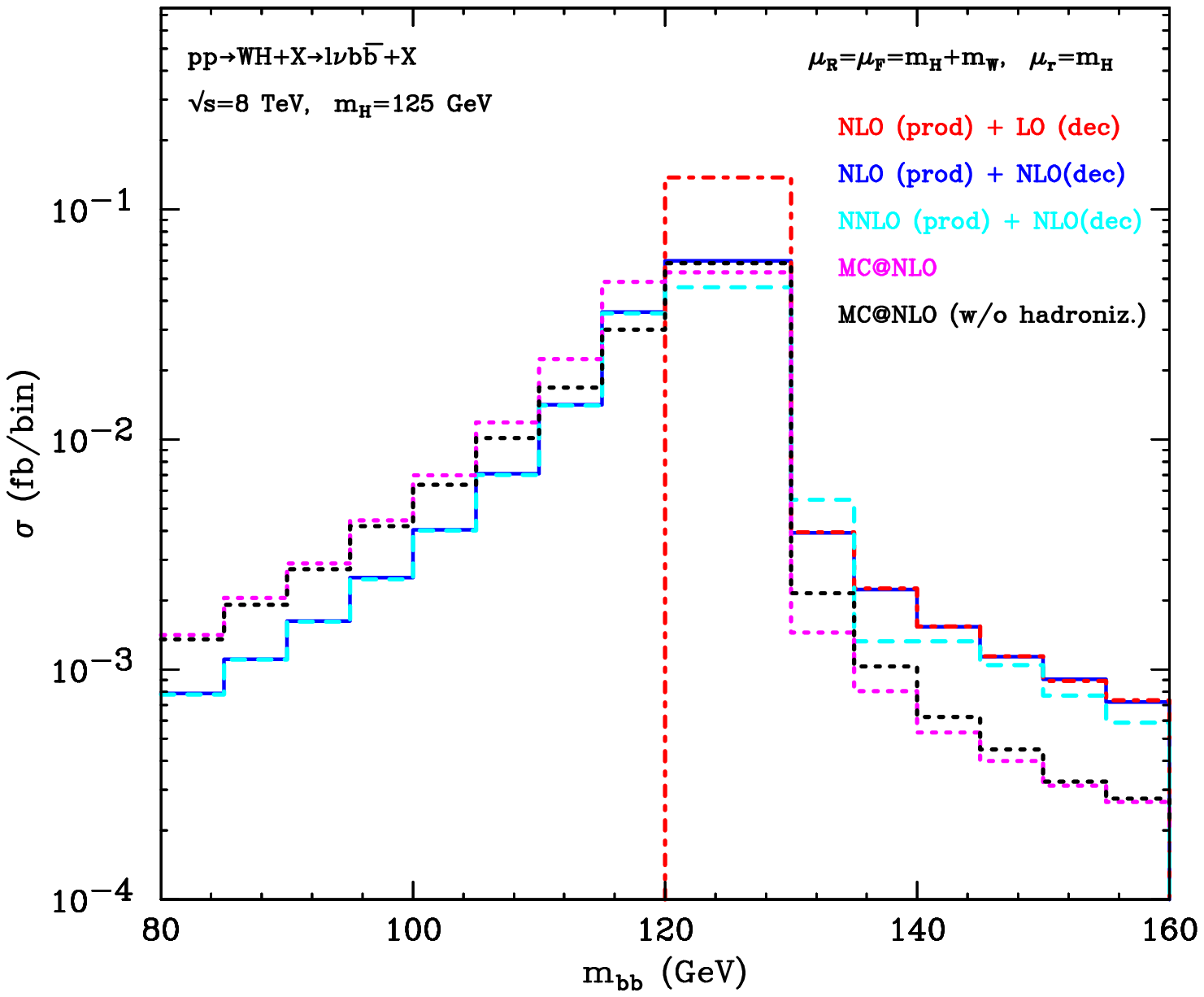}
\end{tabular}
\end{center}
\caption{\label{fig:mbb}
{\em Invariant mass distribution of the pair of $b$-jets
computed at
NLO with LO decay (red dot-dashes), NLO with NLO decay (blue solid), NNLO with NLO decay (cyan dashes), \MC@NLO without hadronization (black dots) and with default \MC@NLO (magenta dots).
The applied cuts are described in the text. Left panel: without jet veto. Right panel: with jet veto.}}
\end{figure}

We finally consider the invariant mass distribution of the pair
of $b$-jets.
In Fig.~\ref{fig:mbb} we plot such distribution  at the various perturbative orders and we compare it with the result obtained
with \MC@NLO. The plot on the left panel corresponds to the case in which the selection cuts discussed above (but no jet veto) are applied; the plot on the right panel
is obtained by further applying the light-jet veto.

We start our discussion by noting that when only NLO corrections to the production are considered (dot-dashes histograms in Fig.~\ref{fig:mbb}), the invariant
mass distribution is kinematically bounded by $m_{bb}\geq m_H$.
This is because the parton radiated from the initial state can
be clustered in one of the two $b$-jets, thus increasing their total invariant mass. If this is not
the case, then we simply have $m_{bb}=m_H$.
Equivalently, if only NLO corrections to the decay are considered (this case is not shown in Fig.~\ref{fig:mbb}), the invariant mass distribution is bounded by $m_{bb}\leq m_H$ because the gluon radiated off the $b{\bar b}$ pair can form a jet on its own, thus decreasing the invariant mass of the dijet system.
As already discussed above,
in such situations
the inclusion of further radiative effects leads to
perturbative instabilities \cite{Catani:1997xc}, which spoil the reliability
of the fixed-order expansion around the boundary and would
require a resummation of the soft-gluon contributions to all orders.
As done in Fig.~\ref{fig:cuts}, we
can restore the validity of the fixed-order prediction
by choosing a wider bin around the boundary region $m_{bb}=125$ GeV,
which is also the peak region in the invariant mass distribution.

The fact that NLO corrections to production and decay act in opposite regions of the invariant mass spectrum allows us to clearly assess their different impact.


In the high-mass region, NLO corrections to the decay are irrelevant,
nevertheless \MC@NLO underestimates the cross section.
Such effect is due to the parton shower: in the NLO calculation,
events in which an initial state parton has been clustered with one of the two $b$-quarks will have $m_{bb}>m_H$, but
the final state radiation from the parton shower
will effectively reduce the dijet invariant mass $m_{bb}$.
In this region the NNLO effect is positive, and is partially washed out when the jet veto is applied.
In the low-mass region the parton shower is more effective than the fixed order calculations in reducing the invariant mass of the dijet system, and the \MC@NLO prediction is higher
than the NLO and NNLO result.
The effect of hadronization on the \MC@NLO result is relatively small: switching off hadronization the difference
between the \MC@NLO result and the NLO and NNLO results is reduced only partially.
In summary, with respect of the \MC@NLO prediction, the effect of higher-order QCD corrections is
to make the invariant mass distribution harder.

\section{Numerical results at $\sqrt{s}=14$ TeV}
\label{sec:lhc14}

\begin{figure}[th]
\begin{center}
\begin{tabular}{cc}
\includegraphics[width=.50\textwidth]{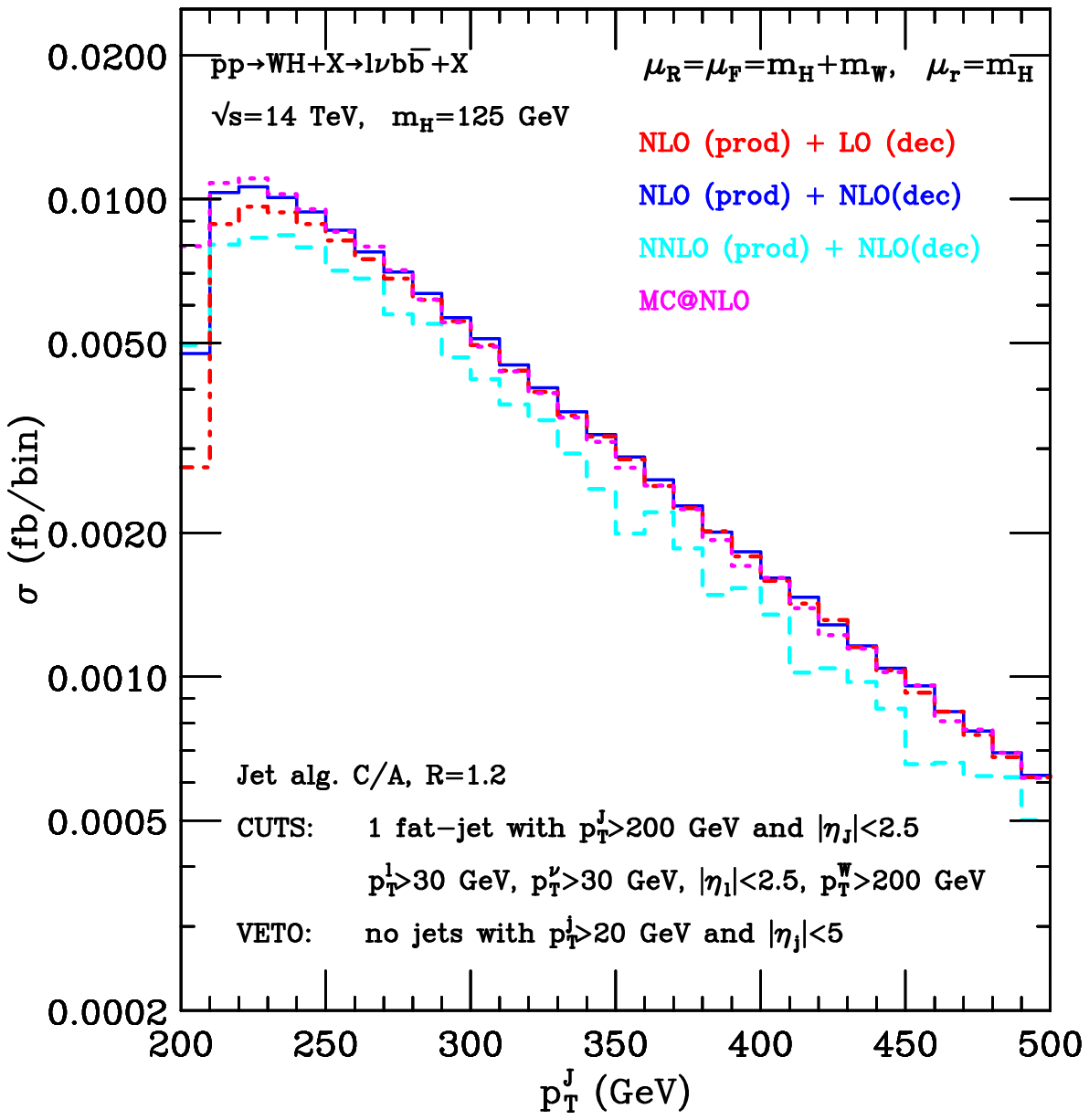}
\includegraphics[width=.45\textwidth]{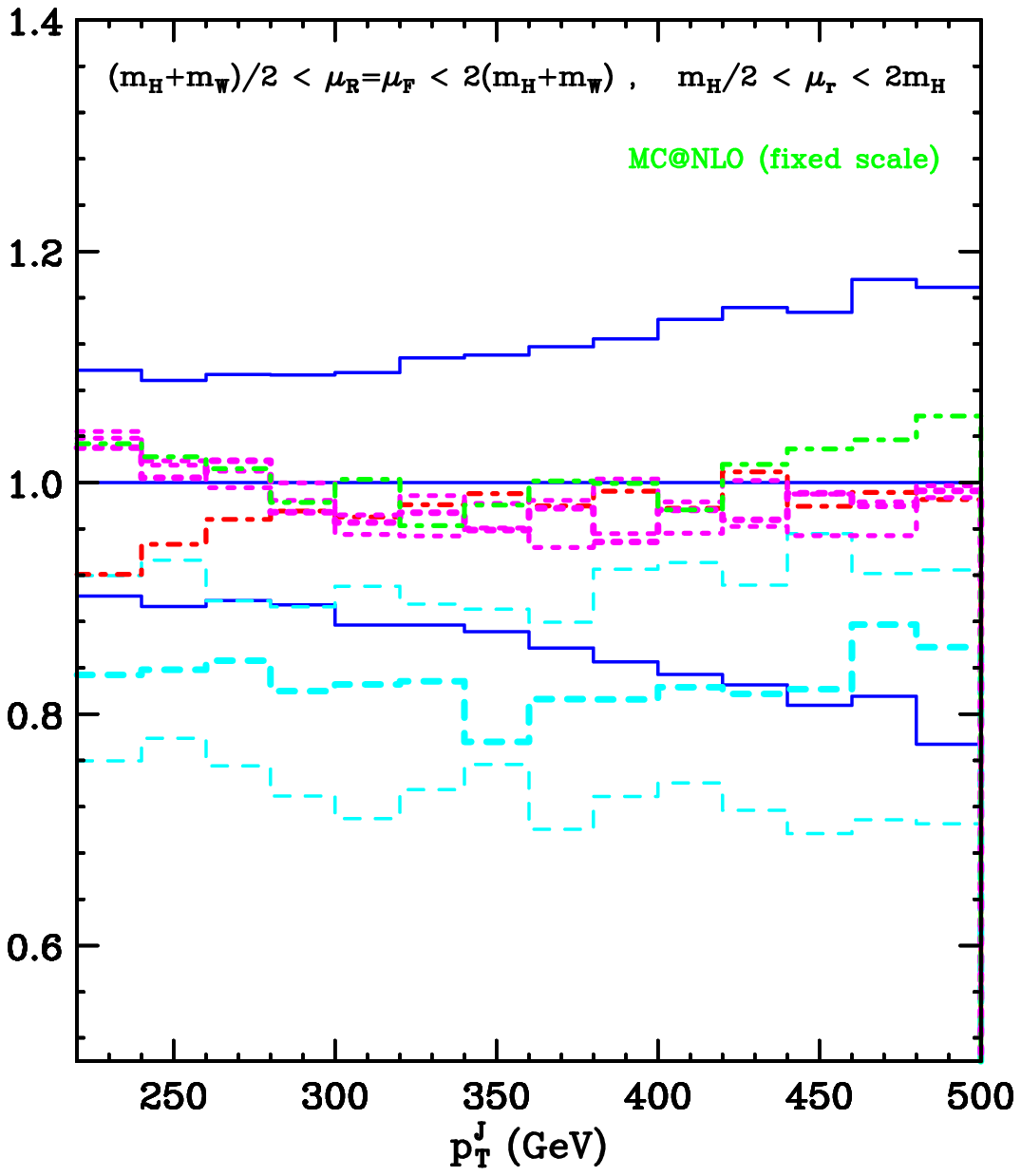}
\end{tabular}
\end{center}
\caption{\label{fig:14tev}
{\em 
Left panel: Transverse-momentum distribution of the fat jet computed at NLO with LO decay (red dot-dashes),
NLO with NLO decay (blue solid), NNLO with NLO decay (cyan dashes) and with \MC@NLO (magenta dots).
Right panel: The same distribution normalized to the full NLO result; the \MC@NLO result (green dots) with fixed scale is also shown. 
The applied cuts are described in the text.}}
\end{figure}

In this Section we consider the case of \WH\ production at the LHC with $\sqrt{s}=14$ TeV.
We follow the selection strategy of Ref.~\cite{Butterworth:2008iy}, that we have already considered in Ref.~\cite{Ferrera:2011bk}.
The Higgs boson is selected at large transverse momenta through its decay
into a collimated $b{\bar b}$ pair.

We require the charged lepton to have $p_T^l > 30$ GeV and $|\eta_l|< 2.5$, and the missing transverse momentum
of the event to fulfil $p_T^{\rm miss}> 30$ GeV.
We also require the $W$ boson to have $p_T^W>200$ GeV.
Jets are reconstructed with the
Cambridge/Aachen algorithm \cite{Dokshitzer:1997in,Wobisch:1998wt}, with $R=1.2$.
One of the jets ({\it fat jet}) must have $p_T^J>200$ GeV\footnote{We note that these {\it symmetric} cuts
on the transverse momenta of the Higgs and the $W$ boson lead to well known perturbative instabilities \cite{Frixione:1997ks,Banfi:2003jj} in
the fixed order predictions around the cut. Here we simply ignore this problem and focus on the $p_T$ distribution of the Higgs candidate sufficiently above  $p_T=200$ GeV.}
and $|\eta_J|<2.5$ and must contain the $b{\bar b}$ pair. In the \MC@NLO simulation, the
fat jet is required to contain two $B$ hadrons.
We also apply a veto on further
light jets with $p^j_T>20$ GeV and $|\eta_j|< 5$.
The corresponding accepted cross sections and uncertainties are reported in Table \ref{table2}.
We see that, compared to the analysis at $\sqrt{s}=8$ TeV, the effect of the jet veto is more important, and it leads to a reduction of
the accepted cross section of about a factor of two for the NLO and \MC@NLO predictions, and
by $57\%$ at NNLO.
This reduction of the accepted cross section with respect to the case in which the jet veto is not applied
is accompanyed by a significant increase in the scale uncertainty in our fixed order results.
The reason for this increased sensitivity is twofold:
first, the typical invariant mass of the \WH\ system
in this case is larger, due to the higher $p_T$ required for both the Higgs and the $W$ boson; second, the typical scale
of QCD radiation is higher, due to the higher centre-of-mass energy, being
the jet veto scale the same used at $\sqrt{s}=8$ TeV.

Our results for the $p_T$ distribution of the Higgs candidate in
this {\it boosted} scenario are reported in Fig.~\ref{fig:14tev}. Comparing with the
results of the previous Section we see clear differences. First of all,
the effect of NLO corrections for the decay is much smaller, and essentially
negligible for $p_T\gtap 300$ GeV. This is not unexpected: in this kind of analysis the (boosted) fat jet is essentially {\it inclusive} over QCD radiation
and the impact of the QCD corrections to the decay is well accounted for by the
inclusive QCD corrected \Hbb\ branching ratio.  This observation is important because it confirms
the validity of the results presented in Ref.~\cite{Ferrera:2011bk}, where the corrections to
the decay were neglected.
The NLO scale uncertainty, obtained as in Sect.~\ref{sec:lhc8}, is about $\pm 10\%$ at $p_T\gtap 200$ GeV,
and it increases to about $\pm 20\%$ at $p_T\sim 500$ GeV.
We also note that the \MC@NLO prediction is in good agreement as well with
the complete NLO result.
In Table~\ref{table2}
we see that, as observed in Sect.~\ref{sec:lhc8},
the \MC@NLO prediction has very small uncertainty, much smaller than the scale
uncertainties of the other calculations: we thus conclude that,
most likely, such uncertainty cannot be considered reliable.
The \MC@NLO result computed with fixed scale is consistent with the \MC@NLO band except in the very high-$p_T$ region.
The NNLO result is smaller than NLO by about $16\%$,
consistently with what shown in Fig.~2 of Ref.~\cite{Ferrera:2011bk},
and it is at the border of the band from scale variations.
The effect is thus qualitatively similar to what discussed in Sect.~\ref{sec:lhc8} but larger in size.
The NNLO scale uncertainty band overlaps with the NLO band, and is smaller in size.

In summary, our results on the boosted scenario at $\sqrt{s}=14$ TeV
show that the shape of the Higgs $p_T$ spectrum is rather stable,
with uncertainties at the few percent level. The normalization of the accepted cross section has instead larger uncertainties with respect to the analysis at $\sqrt{s}=8$ TeV. From Fig.~\ref{fig:14tev} we estimate
that these uncertainties are at the $10-15\%$ level.
An alternative way to estimate the perturbative uncertainty could be to follow the prescription of Ref.~\cite{Stewart:2011cf}. A reduction of the uncertainty can be obtained by performing the resummation of
the large logarithmic contributions, along the lines of Refs.~\cite{Banfi:2012jm,Becher:2013xia,Stewart:2013faa}.
Note, however, that this would be possible at present only by neglecting the radiation from the $b{\bar b}$ pair,
whose effect, however, is marginal in the boosted scenario.

\begin{table}[h]
\begin{center}
\begin{tabular}{|c|c|c|c|c|}
\hline
$\sigma$ (fb) & NLO (with LO dec.) & NLO (full) & NNLO (with NLO dec.) & \MC@NLO \\
\hline
w/o jet veto & $2.54^{+1\%}_{-1\%}$ & $2.63^{+1\%}_{-1\%}$ & $2.52^{+2\%}_{-2\%}$ & $2.82^{+1\%}_{-1\%}$ \\
\hline
w jet veto  & $1.22^{+11\%}_{-14\%}$ & $1.29^{+12\%}_{-13\%}$ &  $1.07^{+8\%}_{-6\%}$ & $1.33^{+1\%}_{-1\%}$ \\
\hline
\end{tabular}
\end{center}
\caption{{\em Cross sections and their scale uncertainties
for $pp\to WH+X\to l\nu b \bar b+X$ at the LHC with $\sqrt{s}=14$ TeV. 
The applied cuts are described in the text.}}
\label{table2}
\end{table}

We now move to consider the invariant mass distribution of the fat jet.
\begin{figure}[th]
\begin{center}
\begin{tabular}{cc}
\includegraphics[width=.8\textwidth]{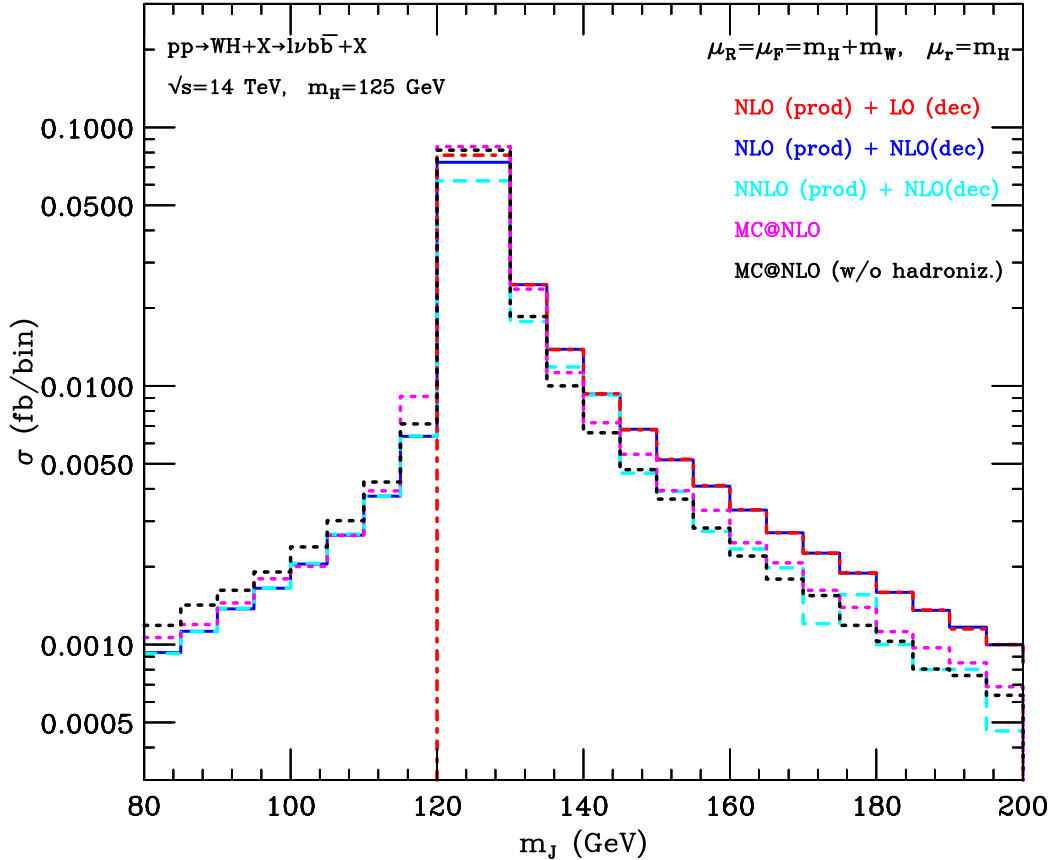}
\end{tabular}
\end{center}
\caption{\label{fig:mbb14}
{\em  Invariant mass distribution of the fat jet computed at NLO with LO decay (red dot-dashes),
NLO with NLO decay (blue solid), NNLO with NLO decay (cyan dashes), \MC@NLO without hadronization (black dots) and with default \MC@NLO (magenta dots).
}}
\end{figure}
In Fig.~\ref{fig:mbb14} we report our fixed-order predictions for this distribution and compare them to the result
from \MC@NLO. We immediately see that, contrary to what happens in Fig.~\ref{fig:mbb},
the invariant mass distribution of the fat jet has a more pronounced tail at high mass.
This somewhat confirms what we have already observed, that QCD radiative effects on the production process, which naturally populate the high-mass region,
are those that are more relevant in the fat-jet analysis.
The fixed order and \MC@NLO results for $m_J<m_H$ are essentially identical,
whereas at $m_J>m_H$ the reduction in the cross section due to the parton shower
is similar in size to the (negative) NNLO effect.
From Fig.~\ref{fig:mbb14} we conclude that, contrary to what happens in the analysis at $\sqrt{s}=8$ TeV (see Sect.~\ref{sec:lhc8}), 
the invariant mass distribution
is relatively stable with respect to radiative corrections.

\section{Summary}
\label{sec:summa}

In this paper we have studied the effect of QCD radiative corrections
on the associated production of the Higgs boson with a $W$ boson in hadronic collisions, 
followed by the $W\to l \nu_l$ and the $H\to b\bar b$ decays. 
We performed a QCD calculation that includes the contributions from higher-order 
radiative corrections up to NNLO for the \WH\ production and up to NLO for the $H\to b \bar b$ decay.
By exploiting the narrow-width approximation (see Eq.~(\ref{master})) and
by appropriately normalizing the $Hbb$ coupling, the prediction we obtain is insensitive to
higher-order corrections to the \Hbb\ decay for a completely inclusive observable.
Having accounted for the fully exclusive \Hbb\ decay at the NLO, our calculation should thus provide a reliable
approximation to the complete NNLO calculation.

Our computation is implemented in a parton level Monte Carlo program
that allows us to apply arbitrary kinematical cuts on the $W$ and $H$ decay products
and on the accompanying QCD radiation. A public version of this program will be available
in the near future.

We have focused our study on the transverse momentum and the invariant mass distributions
of the Higgs candidate, which are the most
relevant observables for the experimental analysis at the LHC.
We have compared the effects of the QCD radiative corrections at various level of accuracy
with the results obtained with the \MC@NLO event generator.

We find that NLO corrections to the \Hbb\ decay can be important to
obtain a reliable $p_T$ spectrum of the Higgs boson, but that, in the cases of interest,
the final state
radiation is well accounted for by the Monte Carlo parton shower.
The jet veto that is usually applied on additional light jets challenges the stability of the perturbative expansion. Nonetheless, with the selection cuts applied in the $\sqrt{s}=8$ TeV analysis,
we have shown that the theoretical prediction of $p_T$ spectrum of the Higgs candidate appears
under good control. The impact of the jet veto is larger in the boosted analysis at $\sqrt{s}=14$ TeV, and
perturbative uncertainties are more sizeable. 
NNLO corrections to the production process
decrease the cross section by an
amount which depend on the detail of the applied cuts, but, in all cases we have considered,
NNLO corrections have a mild effect on the shape of the Higgs $p_T$ spectrum.

The effect of higher-order QCD radiative corrections on the invariant mass distribution
of the Higgs candidate is different in the $\sqrt{s}=8$ and 14 TeV analyses.
In the $\sqrt{s}=8$ TeV analysis the higher-order QCD effects tend to make the invariant mass distribution harder
with respect to the \MC@NLO prediction. In the fat-jet analysis at $\sqrt{s}=14$ TeV, the invariant mass of the fat
jet is rather stable when higher-order QCD effects are considered.

We finally
point out that a possible continuation of the study presented here could be along the lines
of Refs.~\cite{Hamilton:2013fea,Luisoni:2013cuh}, to perform a full NNLO+PS simulation for this process.

\section*{Acknowledgments}

We would like to thank Stefano Catani, Niklas Mohr and Andrea Rizzi for helpful discussions,
Stefano Frixione and Bryan Webber for their help with the \MC@NLO event generator,
and Andrea Banfi for comments on the manuscript.
This research was supported in part by 
the Research Executive Agency (REA) of the European Union
under the Grant Agreement number PITN-GA-2010-264564 ({\it LHCPhenoNet}).
The work of FT is partially supported
by MIUR under project 2010YJ2NYW.

\bibliographystyle{mystyle}
\bibliography{whmc_rev}

\end{document}